\documentclass[apj]{emulateapj}

\pdfoutput=1
\usepackage{hyperref}
\usepackage{amsmath,amstext}
\usepackage[T1]{fontenc}
\usepackage[figure,figure*]{hypcap}
\usepackage{natbib}
\usepackage{enumitem}

\shorttitle{H$_2$-H\lowercase{e} CIA in cool white dwarfs atmospheres}
\shortauthors{Blouin, Kowalski \& Dufour}

\begin{document}

\submitted{Accepted for publication in The Astrophysical Journal}

\title{Pressure distortion of the H$_2$-H\lowercase{e} Collision-Induced Absorption at the photosphere of Cool White Dwarf Stars}

\author{S. Blouin\altaffilmark{1,2}}
\author{P.M. Kowalski\altaffilmark{2}}
\author{P. Dufour\altaffilmark{1}}

\altaffiltext{1}{D\'epartement de Physique, Universit\'e de Montr\'eal, Montr\'eal, QC H3C 3J7, Canada; sblouin@astro.umontreal.ca, dufourpa@astro.umontreal.ca.}
\altaffiltext{2}{IEK-6 Institute of Energy and Climate Research, Forschungszentrum J\"ulich, 52425 J\"ulich, Germany; p.kowalski@fz-juelich.de.}

\begin{abstract}
Collision-induced absorption (CIA) from molecular hydrogen is a dominant opacity source in the atmosphere of cool white dwarfs. It results in a significant flux depletion in the near-IR and IR parts of their spectra. Because of the extreme conditions of helium-rich atmospheres (where the density can be as high as a few g/cm$^3$), this opacity source is expected to undergo strong pressure distortion and the currently used opacities have not been validated at such extreme conditions. To check the distortion of the CIA opacity we applied state-of-the-art ab initio methods of computational quantum chemistry to simulate the CIA opacity at high densities. The results show that the CIA profiles are significantly distorted above densities of $0.1\,{\rm g/cm}^3$ in a way that is not captured by the existing models. The roto-translational band is enhanced and shifted to higher frequencies as an effect of the decrease of the interatomic separation of the H$_2$ molecule. The vibrational band is blueward shifted and split into $Q_R$ and $Q_P$ branches, separated by a pronounced interference dip. Its intensity is also substantially reduced. The distortions result in a shift of the maximum of the absorption from $2.3\,\mu{\rm m}$ to $3-7 \mu{\rm m}$, which could potentially explain the spectra of some very cool, helium-rich white dwarfs.
\end{abstract}

\section{Introduction}

White dwarf stars represent the last stage in the evolution of most stars. Deprived of any internal energy source, they slowly cool down during billions of years. Since their cooling rate and cooling time can be accurately calculated, these stars are excellent cosmochronometers \citep{fontaine2001potential,winget1987independent}. The analysis of a large sample of white dwarf stars can reveal the age and the historical stellar formation rate of various stellar populations \citep{bergeron1997chemical,bergeron2001photometric,hansen2002white,tremblay2014white,winget1987independent}.

To extract the cooling time of a white dwarf star from its spectrum, it is necessary to accurately estimate its atmospheric parameters, namely the effective temperature $T_{\mathrm{eff}}$, the surface gravity $\log g$ and the chemical composition. To do so, detailed atmosphere models are used to fit the observed spectral energy distribution (SED) \citep[e.g.,][]{bergeron1997chemical,bergeron2001photometric}. While current atmosphere models are able to successfully reproduce the spectra of most white dwarf stars, cool ($T_{\mathrm{eff}} < 6000$\,K) helium-rich objects represent a challenge. This is because at low temperatures helium becomes increasingly transparent, allowing the photosphere of cool helium-rich white dwarfs to reach fluid-like densities \citep[up to a few~g/cm$^3$, ][]{bergeron1995new,kowalski2010understanding}. Under such conditions, the average interatomic distance is very short ($\sim 2 \,$\AA) and the collective interactions between particles are strong, affecting the chemistry and physics. Various studies have proposed non-ideal corrections to the equation of state \citep{bergeron1995new,FGV77,SJ98}, the chemical abundances of species \citep{kowalski2006dissociation,kowalski2007equation}, the opacities \citep{iglesias2002density,kowalski2006found,kowalski2014infrared}, and the radiative transfer \citep{kowalski2004radiative} in dense helium-rich media. Despite these improvements, atmosphere models have problems to correctly reproduce the spectra of cool, helium-rich stars, and thus fail to deliver reliable atmospheric parameters for many of them. The reproduction of the near-IR and IR parts of the spectra is especially problematic \citep{bergeron2002model,gianninas2015ultracool}. This points to a problem with the current description of the CIA opacity from molecular hydrogen as a trace species in a dense helium-rich medium \citep[see the case of LHS 3250,][]{kilic2009spitzer,kowalski2013cool}.

The importance of CIA opacity from molecular hydrogen in the atmospheres of cool white dwarfs was discussed for the first time by \citet{ML78}. This absorption results from collisions of H$_2$, which alone is IR inactive, with other particles \citep[H, H$_2$, He,][]{frommhold1993collision}, leading to the induction of an electric dipole moment through the formation of a super-molecular complex \citep[e.g., H$_2$He,][]{lenzuni1991rosseland}. This absorption mechanism has been observed experimentally \citep{birnbaum1978far,birnbaum1987experimental,bouanich1990collision,brodbeck1995collison,CMC52,chisholm1954induced,crawford1950infra,hare1958pressure} and computed by various authors \citep{abel2013collision,abel2012infrared,borysow1992new,B02,BF89,BTF85,BFM89,BJZ97,BJF01,frommhold1993collision,GF01,jorgensen2000atmospheres,lenzuni1991rosseland}. Its importance in the modeling of stellar atmospheres, including white dwarfs, was highlighted by \citet{lenzuni1991rosseland} and its implementation in the context of white dwarf atmosphere modeling was discussed in follow-up studies \citep[e.g.,][]{bergeron1995new,SJ98}. We note however that most of these studies focus on modeling binary H$_2$-perturber collisions. The density correction that accounts for the three-body collisions is applied on top of these calculations \citep[e.g.,][]{lenzuni1991rosseland,SJ98} and comes from the experimental measurements of \citet{hare1958pressure}. Because of the limitations of these static-like calculations, the pressure-induced distortion of the roto-translational and vibrational bands, which arises from the kinetics of the collision, have not been considered.

In this study we applied a state-of-the-art ab initio molecular dynamics method to directly simulate the H$_2$-He CIA absorption in a density regime where many-body collisions are important ($\rho>0.1 \,{\rm g/cm}^3$, $T>\rm 1000\,K$). This allows us for the first time to observe the pressure distortion of the absorption profiles at such extreme and previously unexplored conditions. Moreover, we constructed a model for this distortion, which can be easily implemented in existing white dwarf atmosphere codes.

In Section \ref{sec:theory}, we provide a brief description of the physics of CIA and the details of our simulation strategy. The discussion of the results of our virtual experiments are presented in \mbox{Section \ref{sec:results}}, where we compare the obtained data with previous calculations and the available experimental data. Here we also provide a detailed analysis of the simulated H$_2$-He CIA distortion that arises at densities above $0.1\,{\rm g/cm}^3$ and an analytical model of the distortion is given in Section \ref{sec:modelcia}. In Section \ref{sec:models}, we discuss the resulting atmosphere models and their implications for our understanding of white dwarf spectra. Our conclusions and future directions are provided in Section \ref{sec:conclusion}.

\section{Theoretical framework}
\label{sec:theory}

\subsection{Collision-induced absorption}
\label{sec:cia}
Due to symmetry, an isolated H$_2$ molecule in its electronic ground state has no electric dipole moment. Hence, it can only absorb photons through less probable electric quadrupole transitions, which makes it effectively IR inactive. However, in a dense medium, the interaction of H$_2$ molecule with surrounding particles (binary collisions at low densities) leads to a distortion of its charge distribution and induction of a (small) electric dipole moment. This induced dipole allows the absorption of photons via electric dipole transitions, which are more probable than the aforementioned quadrupole transitions. This phenomenon is known as collision-induced absorption \citep{frommhold1993collision}.

\begin{figure}
\centering
\includegraphics[width=\columnwidth]{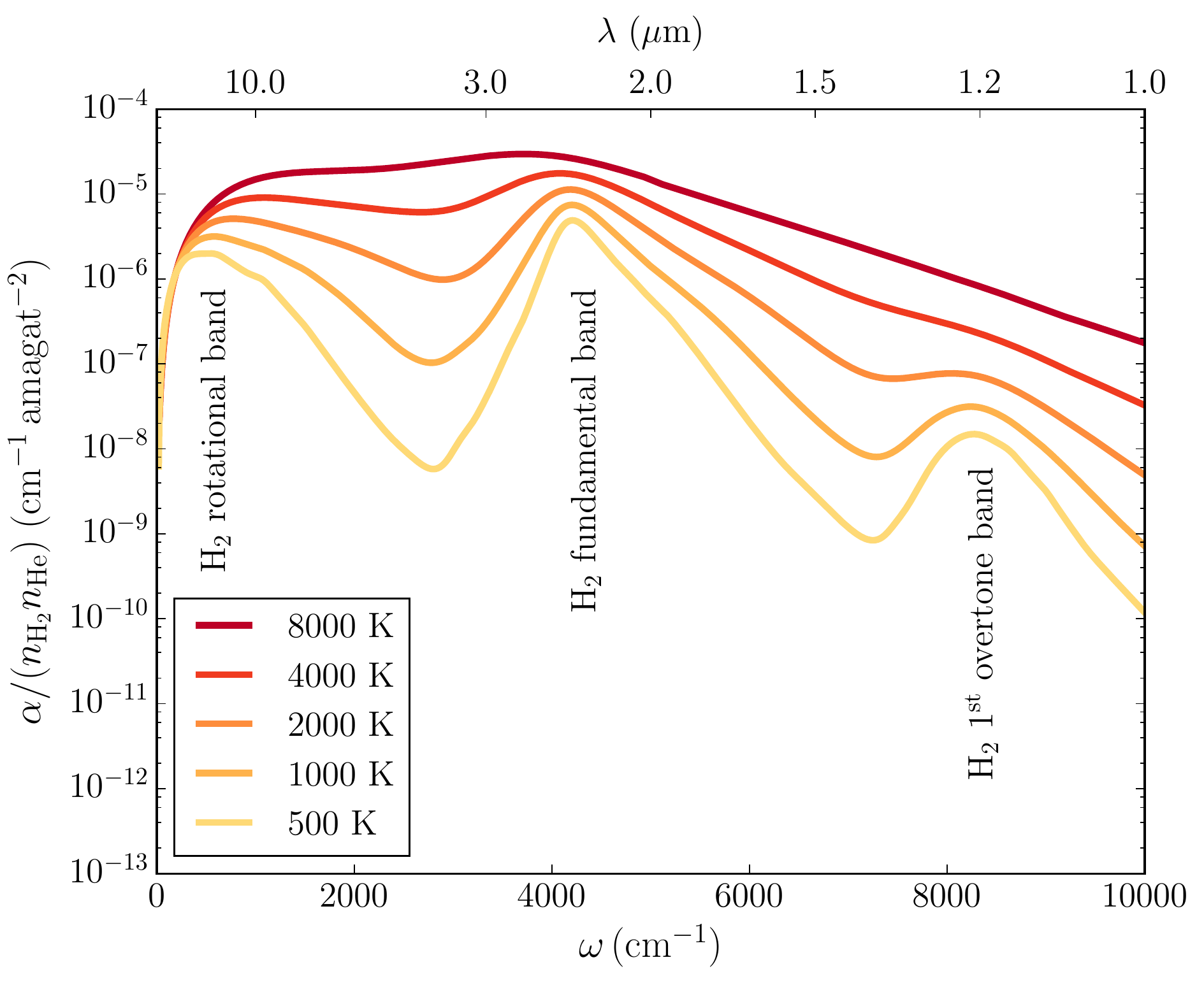}
\caption{H$_2$-He CIA spectra for different temperatures, as computed by \cite{abel2012infrared}. All spectra are divided by the number density of H$_2$ and He.}
\label{fig:CIAabel}
\end{figure}

The CIA spectrum of a molecule can be viewed as the sum of two contributions: one from the collisional complex and one from the unperturbed molecule \citep{frommhold1993collision}. As shown in Figure \ref{fig:CIAabel}, for the H$_2$-He CIA spectrum, the dominant contribution to the spectrum comes from the rovibrational transitions of the unperturbed H$_2$ molecule (which are dipole-forbidden for an isolated H$_2$ molecule). The energy of an absorbed photon, $\hbar \omega$, satisfies the relationship \citep{abel2013collision},
\begin{equation}
E_{\nu,J} + E_r + \hbar \omega = E_{\nu',J'} + E_r',
\end{equation}
where $E_{\nu,J}$ and $E_{\nu',J'}$ are the rovibrational energies of the H$_2$ molecule ($\nu$ and $J$ are the vibrational and rotational quantum numbers, respectively) and $E_r$ and $E_r'$ are the energies of relative motion before and after the interaction. Figure \ref{fig:CIAabel} also shows that the bands arising from rovibrational transitions are broadened at high temperature. This is a consequence of faster collisions (shorter interaction times) at higher temperatures \citep{abel2013collision,lenzuni1991rosseland}. 

To compute CIA spectra, previous investigators \citep[e.g.,][]{abel2012infrared,birnbaum1984theory,borysow1992new,hammer1999emission,meyer1986collision} have relied on an approach combining quantum chemical computations with molecular scattering theory. In such calculations ab initio methods (e.g., M{\o}ller-Plesset calculations, coupled-cluster calculations) are used to obtain accurate potential energy (PES) and induced dipole (IDS) surfaces for the H$_2$-perturber super-molecular complex in the infinite-dilution limit. Then, the PES and IDS are used as inputs in the molecular scattering theory to compute the resulting spectrum. This approach has been very successful in predicting the measured H$_2$-H$_2$ \citep{borysow1992new,BBF00} and H$_2$-He \citep{BFB88,borysow1992new} CIA in the dilute limit.

The three-body collision effects at higher densities are modeled with a $1+\beta\rho$ scaling factor (assuming that binary collisions are proportional to $\rho^2$ and triple collisions to $\rho^3$). We note that this approach has not been validated for densities larger than $\rho=0.26\,{\rm g/cm}^3$ -- the highest density measured by \citet{hare1958pressure} -- and temperatures higher than room temperature. It is also expected to break down for the much higher, fluid-like densities encountered at the photosphere of cool white dwarf stars with helium-rich atmospheres. This is because PES and IDS obtained in the infinite-dilution limit are not expected to correctly capture the distortion of the charge distribution resulting from simultaneous, multi-atomic interactions. In order to check how these interactions affect the CIA spectrum, here we applied the ab initio molecular dynamics method to simulate the H$_2$-He CIA at these extreme helium densities. These simulations can be seen as a virtual experiment that intrinsically accounts for all the many-body effects.

\subsection{Ab initio molecular dynamics simulations}
\label{sec:comput}

We simulated the interactions between H$_2$ and He atoms and the resulting IR absorption spectrum using the same procedure and computational setup as \cite{kowalski2014infrared} in his simulations of He-He-He CIA. We used ab initio Born-Oppenheimer molecular dynamics (MD) method with density functional theory (DFT) to calculate the structure of the H$_2$-He atomic fluid. In this simulation framework, the electronic charge density is recomputed at each simulated time point applying DFT, and the atomic dynamics is driven by the resulting forces and classical dynamics. Hence, contrarily to the static molecular scattering approach, PES and IDS are computed on the fly for the atomic configurations representing real atomic arrangements and dynamics in a dense fluid. This is what makes this approach very suitable for the high-density regime, where multi-particles interactions become important.

Simulations were performed using the CPMD\footnote{\url{http://cpmd.org}} plane-wave DFT code \citep{marx2000ab} with the PBE exchange-correlation functional \citep{perdew1996generalized} and ultrasoft pseudopotentials \citep{vanderbilt1990soft}. To assure the convergence of the electronic charge density we applied a plane-wave energy cut-off of 340\,eV. In each simulation, the cubic supercell that represents the simulation box contained one H$_2$ molecule and either 31, 63 or 127 He atoms. The simulation box length was adjusted to obtain the desired density. Numerous simulations were performed for different $(T,\rho)$ conditions.

At each simulation time step, with a time interval of $0.5\rm \,fs$, the total dipole moment $\mathbf{M}$ of the simulation supercell (i.e., the dipole moment resulting from the total electronic charge density and the distribution of all nuclei in the simulation supercell) was computed using the localized Wannier function approach \citep{berghold2000general,silvestrelli1998maximally}. The IR absorption coefficient $\alpha(\omega)$ was then obtained through the Fourier transform of the dipole moment time autocorrelation function as \citep{silvestrelli1997ab},
\begin{equation}
\alpha (\omega) = \frac{2 \pi \omega^2}{3 c k_B T V} \int_{-\infty}^{\infty} \mathrm{d} t \exp(-i \omega t) \langle \mathbf{M} (t) \cdot  \mathbf{M}(0) \rangle,
\label{eq:alpha}
\end{equation}
where $\omega$ is the wavenumber, $c$ is the speed of light in the fluid, $k_B$ is the Boltzmann constant, $T$ is the temperature, $V$ is the supercell volume and the angle brackets denote the time-autocorrelation function, i.e. $\langle \mathbf{M} (\tau) \cdot \mathbf{M} (0) \rangle = \frac{1}{t_{\rm sim}} \int_0^{t_{\rm sim}} \mathbf{M} (t) \cdot \mathbf{M} (t + \tau) \mathrm{d} t$, with $t_{\rm sim}$ being the total simulation time. For the physical conditions considered in this work, the dipole moment induced on the hydrogen molecule is larger by a factor of $\approx 100$ than the total dipole moment induced on helium atoms. Therefore, the interactions between helium atoms have a negligible contribution to the simulated infrared absorption spectrum (it is about four orders of magnitude smaller than H$_2$-He CIA, see \citealt{kowalski2014infrared}). Since in cool, helium-rich white dwarf stars the index of refraction $n(\omega)$ departs significantly from unity \citep{kowalski2004radiative}, the actual absorption coefficient that has to be implemented in atmosphere codes is not $\alpha(\omega)$, but $\alpha(\omega)/n(\omega)$. For helium in the atmosphere of white dwarf stars, $n(\omega)$ can be approximated using for instance the semiempirical virial expansion given by \cite{kowalski2007equation}, which accounts for binary and triple interatomic collisions and is consistent with quantum molecular dynamics data for helium densities up to 2\,g/cm$^3$. However, we note that its applicability at higher densities may be inappropriate due to the expected contributions from quadruple and higher-order collisions.

The outlined procedure for the simulation of IR absorption spectra using a molecular dynamics approach and the time evolution of the dipole moment is a well-established method that has been successfully applied in various previous studies. For instance, this technique was used to simulate the IR spectrum of water \citep{guillot1991molecular,iftimie2005decomposing,silvestrelli1997ab}, the absorption spectra of minerals \citep{pagliai2008anharmonic,pagliai2011spectroscopic} and the He-He-He CIA in dense helium for the conditions of white dwarf atmospheres \citep{kowalski2014infrared}.

\subsection{Quality of the computed dipole moments}
\label{sec:dipole}

\begin{figure}
\centering
\includegraphics[width=\columnwidth]{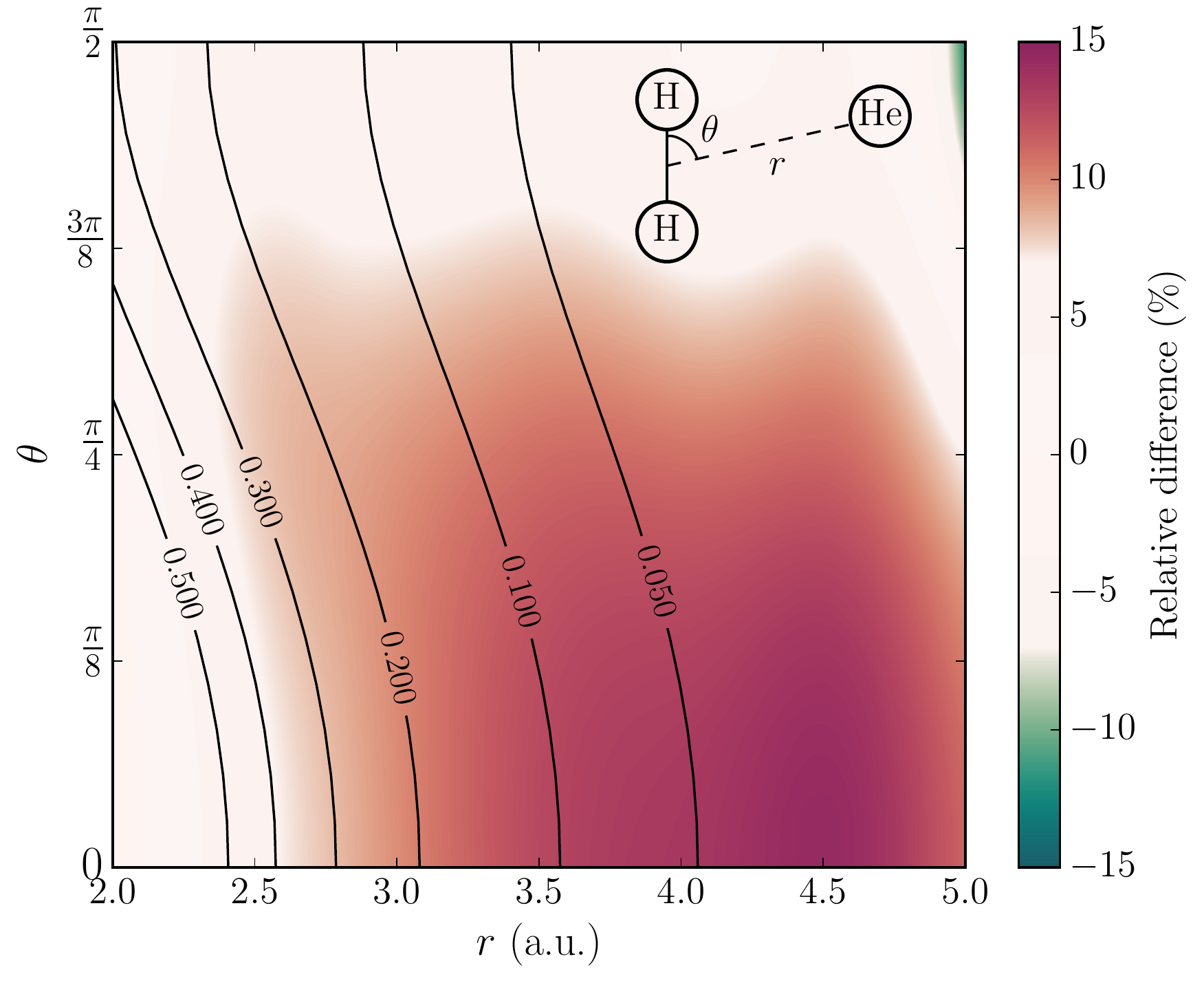}
\caption{Colors indicate the relative difference between the dipole moments of the H$_2$-He complex computed using DFT and those reported by \cite{li2012interaction} and contour lines show the amplitude of the dipole moment in atomic units. As shown in the upper-right corner, $\theta$ is the angle between the H$_2$ bond and the line connecting the center of the H$_2$ molecule to the He atom (the length of this line being $r$). The separation between the two H atoms is fixed at 1.449\,a.u. to allow comparison with the data provided in \cite{li2012interaction}.}
\label{fig:comp_dipole}
\end{figure}

It is well known that due to intrinsic approximations, DFT methods do not result in precise values of dipole moments. Since accurate dipole moments are essential to a correct determination of CIA opacities, we validated the dipole moments of the H$_2$-He super-molecular complex computed with our DFT approach against those computed by \cite{li2012interaction}. The latter study was performed using the couple-cluster method with large basis sets and thus represents the most accurate estimate of dipole moments for this complex. As shown in Figure \ref{fig:comp_dipole}, the agreement between both calculations is satisfactory, as the relative difference is not larger than 15\%. Figure \ref{fig:comp_dipole} also indicates that the DFT dipole moments are systematically slightly overestimated. Interestingly, previous studies obtained similar trends for liquid water. Using DFT and maximally-localized Wannier functions to compute the dipole moment per molecule in liquid water, \cite{silvestrelli1999water} found an average value of 3.0\,D (1.2\,a.u.), which is slightly larger than the coupled-cluster result \citep[$\approx 2.7$\,D,][]{kongsted2002dipole}. Nevertheless, the reasonably good match of the results of our simulations of H$_2$-He CIA in the dilute limit to those obtained by more accurate techniques (as will be discussed in Section 3) shows that these differences do not significantly affect the results of our study.

\subsection{Simulation box size effect}
As our simulations employ periodic boundary conditions to mimic the continuity of the fluid, the simulation box must be large enough to eliminate any side effects resulting from artificial periodicity. Thus, for a given helium density we performed simulations with different numbers of helium atoms per simulation box and different box sizes. We then compared the resulting spectra and found that as long as the cubic simulation box has a dimension of at least $a=10$\,a.u. ($5.3\,$\AA) and contains at least $N=32$ atoms, the resulting spectra are virtually identical. All the results reported here are therefore obtained from simulations that satisfy these criteria.

\subsection{Simulation convergence}

\begin{figure}
\centering
\includegraphics[width=\columnwidth]{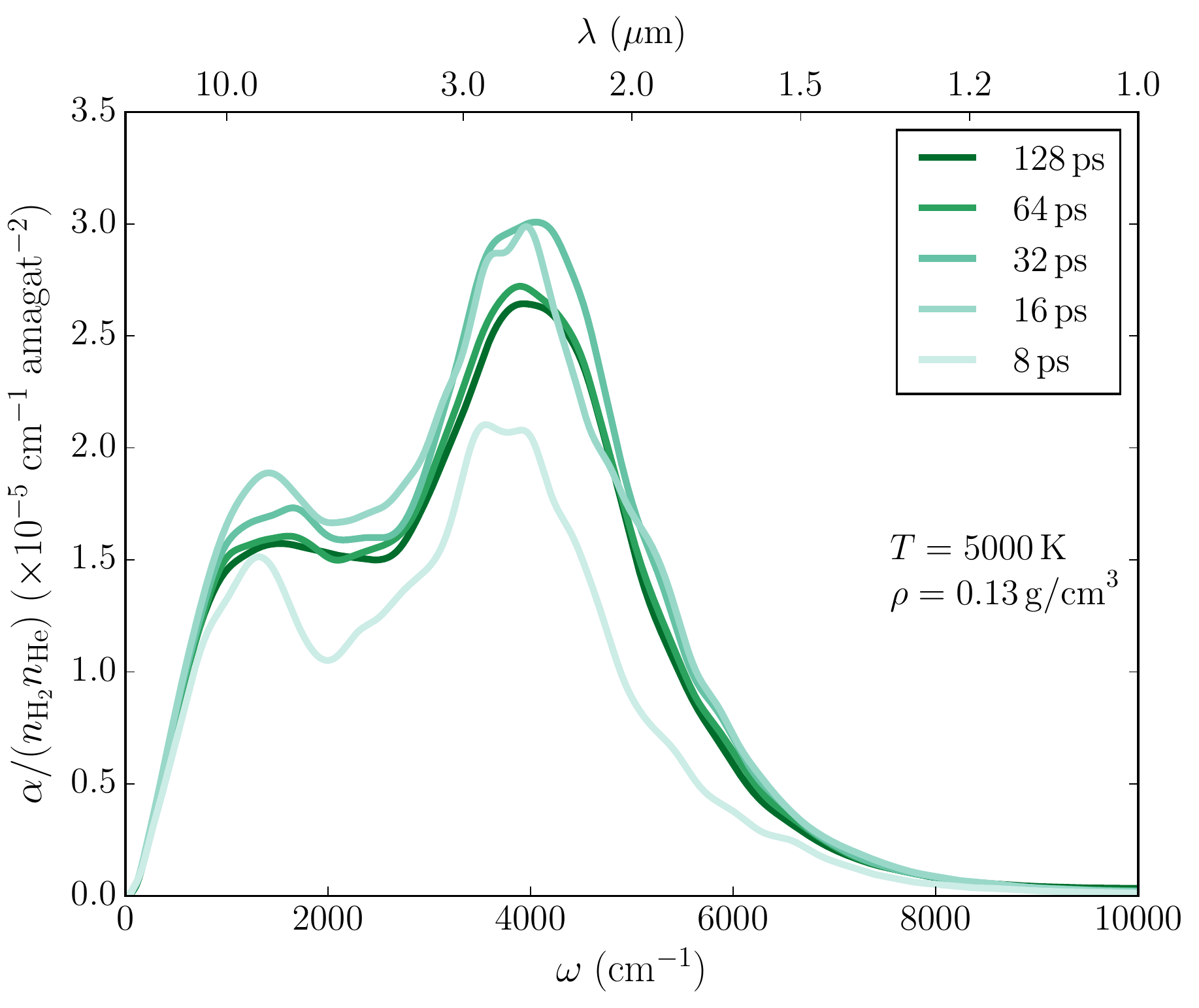}
\caption{
H$_2$-He CIA spectra computed with our DFT-MD simulations for $T=5000\,{\rm K}$ and $\rho=0.13\,{\rm g/cm}^3$. The spectra shown here were obtained using various trajectory lengths, as indicated in the upper-right corner.}
\label{fig:convergence}
\end{figure}

To test the convergence of our simulations in regard to the molecular dynamics simulation time, we computed IR spectra for different trajectory lengths. We found that a 64\,ps trajectory is long enough to attain a satisfactorily good convergence. This is shown in Figure \ref{fig:convergence}, where it can be seen that the 64\,ps and 128\,ps trajectories yield almost identical spectra. Therefore, for all simulations reported here, we use a simulation time of 64\,ps.

\subsection{Quantum effects on the motion of nuclei}
\label{sec:quantum}

Ions in ab initio Born-Oppenheimer molecular dynamics simulations obey the classical laws of motion. Hence, the simulated time evolution of dipole moments follows the classical motion of nuclei. A priori, this is a problematic situation. Our simulations include light atoms, and for $\omega \gtrsim kT / \hbar$, quantum effects are expected to become important. Contrarily to what is done in classical molecular dynamics, one should ideally treat nuclei quantum-mechanically (for instance by applying path integral molecular dynamics). However, when combined with ab initio force determination, this approach becomes computationally very costly. As an alternative, quantum correction factors are applied a posteriori to the simulated "\textit{classical}" IR spectrum, $\alpha_{\rm cl}$. Several different quantum correction factors are discussed in the literature. \cite{ramirez2004quantum} compared many of them and found that the harmonic approximation,
\begin{equation}
\alpha_{\rm corr}=\alpha_{\rm cl}\cdot \frac{\beta \hbar \omega}{1-\exp(-\beta \hbar \omega)},
\end{equation}
performs generally better than the other corrections. Hence, we use this approach to correct our IR spectra for quantum effects. This correction factor is already included in Equation \ref{eq:alpha}.

\section{Results and Discussion}
\label{sec:results}

\subsection{${\rm H}_2$-${\rm He}$ CIA at low density: comparison to previous studies}
\label{sec:previous}

\begin{figure}
\centering
\includegraphics[width=\columnwidth]{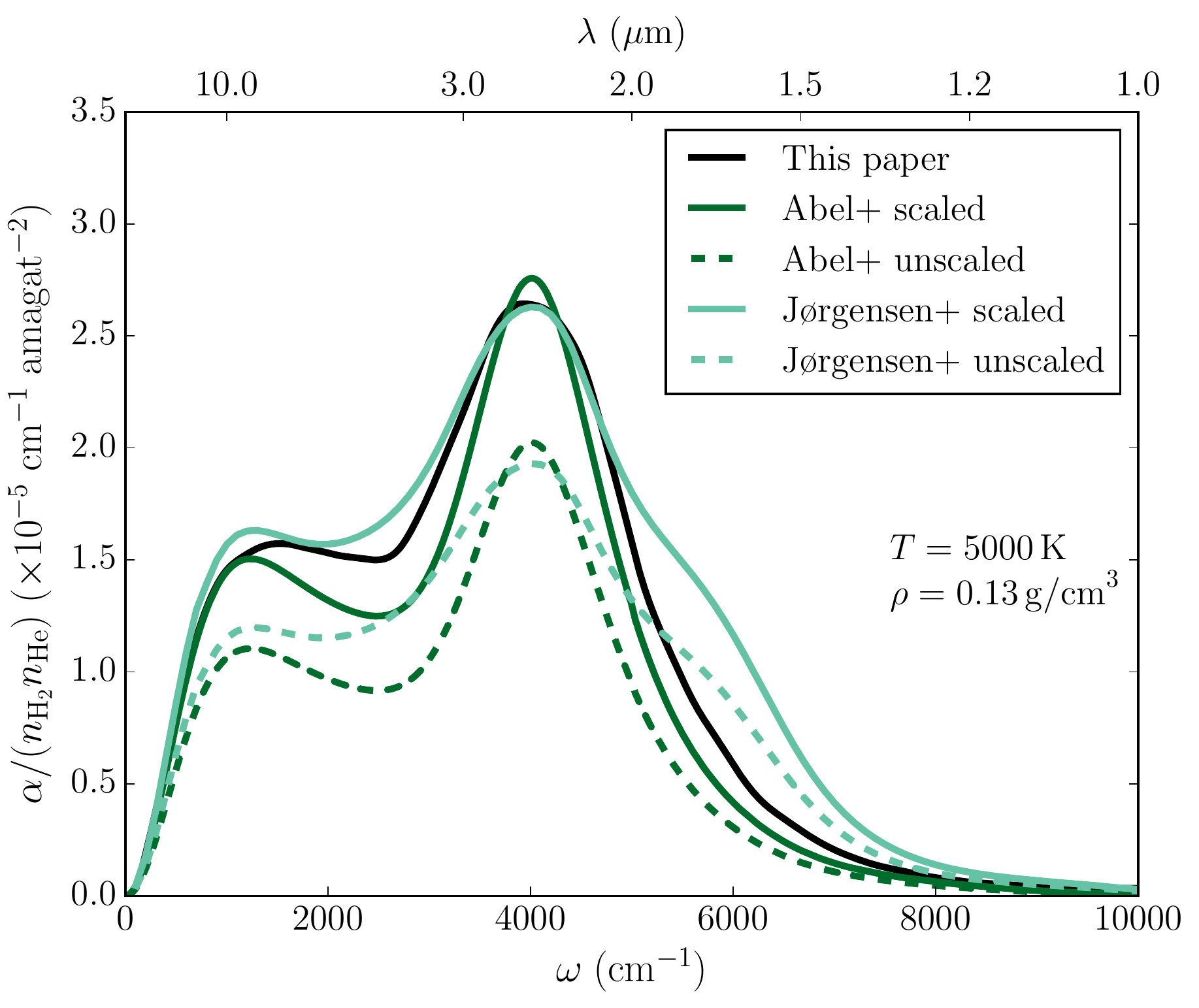}
\caption{The black line is the H$_2$-He CIA spectrum computed with our DFT-MD method. The colored spectra are from \cite{abel2012infrared} and \cite{jorgensen2000atmospheres}. For these two references, the spectra are shown without a three-body correction (dashed lines) and with the density correction of \citealt{hare1958pressure} (solid lines, see Section \ref{sec:previous}).}
\label{fig:convergence2}
\end{figure}

As a first step, we discuss the simulation results at low-density ($\rho =0.13\,{\rm g/cm}^3$) to check if they are consistent with previous calculations. We note that $\rho =0.13\,{\rm g/cm}^3$ is the lowest density we consider, since simulations at $\rho \ll 0.1\,{\rm g/cm}^3$ require much larger simulation cells that are prohibitively computationally intensive. In Figure \ref{fig:convergence2} we compare the simulated H$_2$-He CIA opacity with the absorption profiles computed for $T=5000\,\rm K$ by \cite{abel2012infrared} and \cite{jorgensen2000atmospheres}, and commonly used in white dwarf atmosphere codes \citep[see respectively][]{kowalski2014infrared,bergeron2002model}. We consider the most recent calculations of \cite{abel2012infrared} as the best reference, since these calculations rely on PES and IDS that are more accurate than the ones used by \citealt{jorgensen2000atmospheres} (the latter study used a smaller basis set for their first-principles calculations). We also note that the absorption profiles of \cite{abel2012infrared} are in excellent agreement with room-temperature measurements \citep{birnbaum1978far,birnbaum1987experimental,brodbeck1995collison}.
 
The direct comparison indicates that our simulations overestimate the CIA absorption by $\approx$\,30\%. However, at 0.13\,g/cm$^3$, many-body interactions are already important and the three-body interaction correction (i.e., the $\rho^3$ term) should be taken into account \citep{van1957theory,lenzuni1992hydrogen}. In other words, the absorption profiles of \cite{abel2012infrared} and \cite{jorgensen2000atmospheres} should be scaled by $(1+\beta\rho)$, where $\beta=2.79\,{\rm cm}^3/{\rm g}$ is a coefficient fitted to the experimental data of \cite{hare1958pressure}. As indicated in Figure \ref{fig:convergence2} our simulation profile matches the rescaled profiles of \cite{abel2012infrared} and \cite{jorgensen2000atmospheres}. Some small differences can be noticed in the region between the roto-translational and the fundamental band, and at frequencies above 5000$\,{\rm cm}^{-1}$. This could reflect the uncertainty level caused by the application of the DFT method (see discussion in Section \ref{sec:dipole}). However, these effects are of minor importance and there is an overall satisfactorily good agreement between the simulated and the previously computed CIA profiles, which validates our simulation approach.

We note that \cite{lenzuni1991rosseland} suggested that the strength of the three-body correction can be estimated by scaling the CIA spectrum as $\rho P$ instead of $\rho^2$. Using the tabulated equation of state of \cite{becker2014ab} to get the pressure as a function of density and temperature, we found that the pressure was 33\% higher than the pressure predicted by the ideal gas law for $T=5000\,$K and $\rho=0.13$\,g/cm$^3$, which results in $\beta=3.30\,{\rm cm}^3/{\rm g}$. This value is similar to the $\beta=2.79\,{\rm cm}^3/{\rm g}$ value measured by \citet{hare1958pressure}.

\subsection{${\rm H}_2$-${\rm He}$ CIA at high density}

\begin{figure*}[!tbp]
\centering
\includegraphics[width=\textwidth]{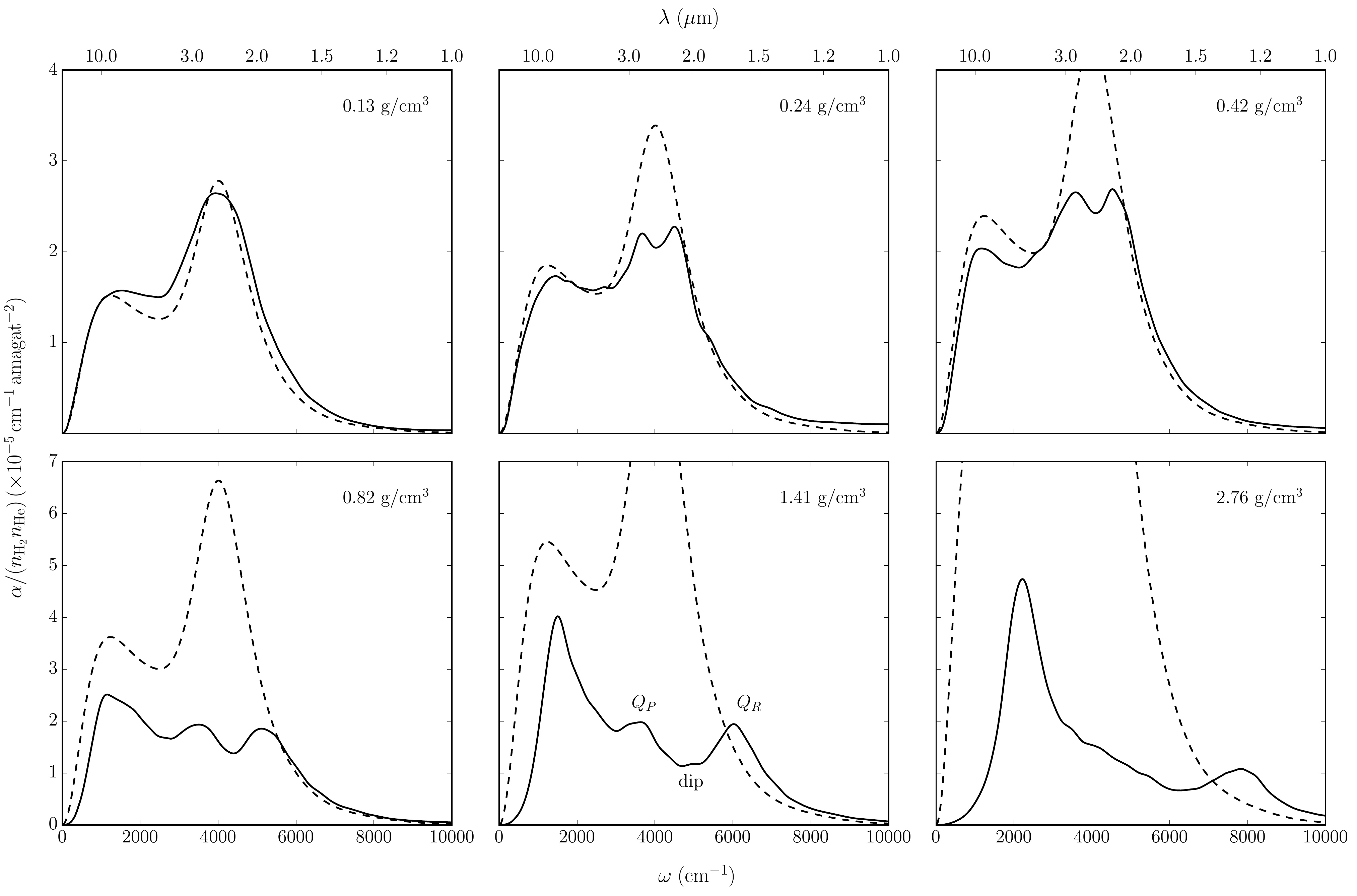}
\caption{The solid lines show the results of our simulations for the absorption coefficient of H$_2$-He CIA as a function of wavenumber, for various densities and $T=5000\,{\rm K}$. The dashed lines are the spectra computed by \cite{abel2012infrared}, scaled by the $1+\beta \rho$ factor discussed in Section \ref{sec:previous}. The scale of the vertical axis is not the same for the top and the bottom plots.}
\label{fig:dens_effect}
\end{figure*}

After demonstrating that our computational approach satisfactorily reproduces the low-density absorption profiles of \cite{abel2012infrared}, we used it to explore how the H$_2$-He CIA profiles behave at higher densities. Figure \ref{fig:dens_effect} shows a density sequence of the simulated CIA profiles and the profiles of \cite{abel2012infrared} rescaled for three-body collisions (Section \ref{sec:previous}). When we compare the two sets of profiles, it is clear that for densities exceeding 0.1\,g/cm$^3$, the spectrum becomes significantly distorted. Three main effects are visible:
\begin{enumerate}[label=(\alph*)]
\item The roto-translational band becomes stronger.
\item The absorption is shifted towards higher frequencies.
\item The fundamental vibrational band is gradually split into two branches with a pronounced, intermediate dip.
\end{enumerate}
Below we discuss these effects in details and try to identify the physical phenomena that cause these distortions.

\subsubsection{(a) Many-body collisions}
\label{sec:many-body}

As already discussed, many-body collisions become increasingly important (compared to binary interactions) under high-density conditions \citep{hare1958pressure,lenzuni1992hydrogen}. This leads to a perturber density-induced scaling of the integrated CIA spectrum,
\begin{equation}
\int \alpha \mathrm{d} \omega = \kappa_2 n_{\mathrm{He}} n_{\mathrm{H}_2} + \kappa_3 n_{\mathrm{He}}^2 n_{\mathrm{H}_2} + \dots,
\label{EQD}
\end{equation}
and the related rescaling of the dilute-limit profiles ($1+\beta \rho$, Section \ref{sec:previous}). As shown in Figure \ref{fig:dens_effect}, the rescaled profiles of \cite{abel2012infrared} match nicely the strength of the simulated roto-translational band up to a density of 0.4\,g/cm$^3$. Therefore, we can confidently conclude that the increase of the roto-translational band occurs as a consequence of three-body collisions. Moreover, our simulation results validate the $1+\beta \rho$ correction of \cite{hare1958pressure} and its implementation in white dwarf atmosphere codes \citep{lenzuni1992hydrogen} for densities up to 0.4\,g/cm$^3$.

\begin{figure}
\centering
\includegraphics[width=\columnwidth]{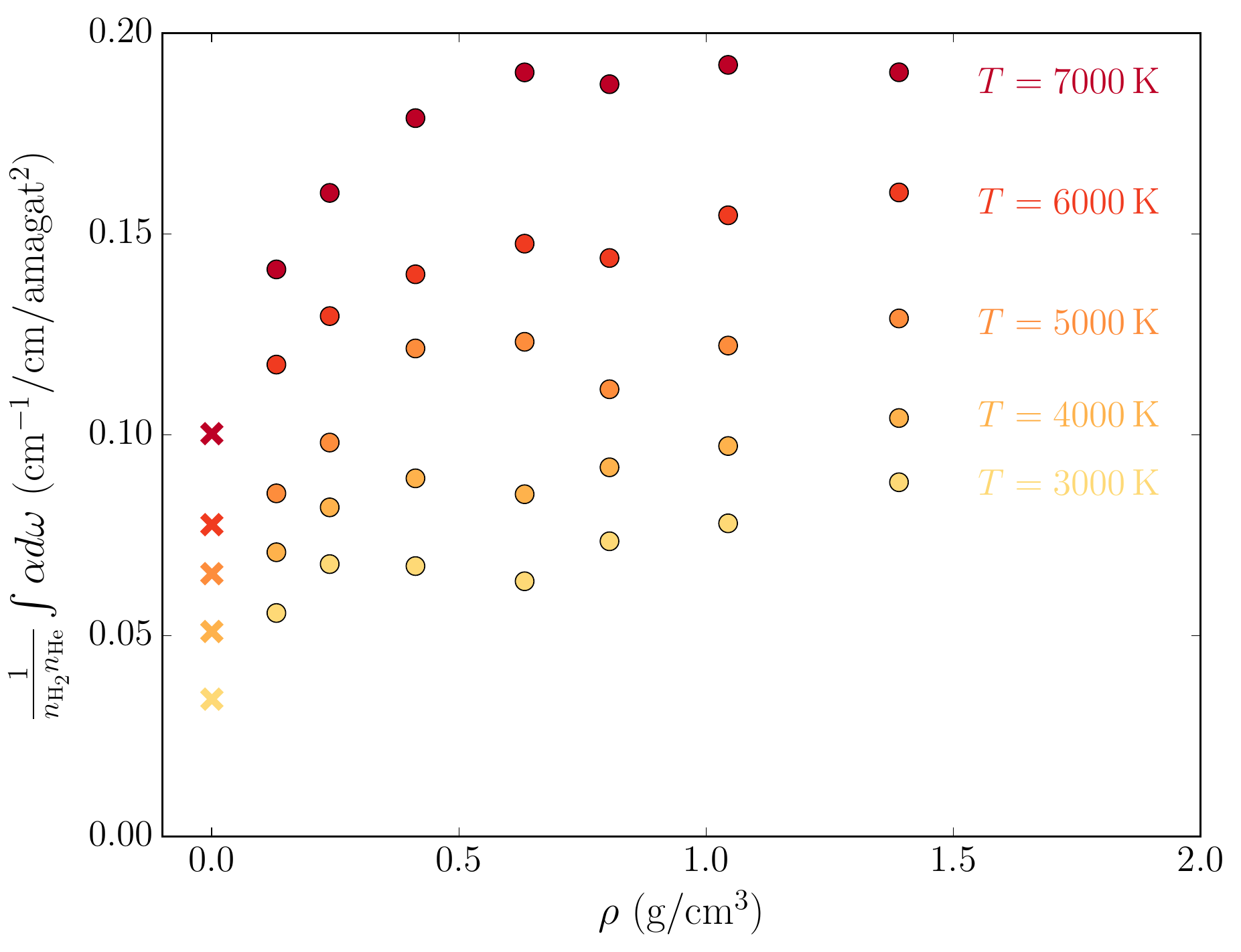}
\caption{Integrated absorption spectra with respect to density. Circles show the results extracted from our ab initio simulations and crosses indicate the values obtained from the low-density calculations of \cite{abel2012infrared}.}
\label{fig:integ}
\end{figure}

Figure \ref{fig:integ} shows how the integrated absorption spectrum changes as a function of density and temperature. It can be seen that the relation between the integrated profile and density is linear up to a density of $\approx 0.5 \,$g/cm$^3$ and thus that it follows Equation \ref{EQD}. These results are consistent with the measurements of \citet{hare1958pressure}, who found a linear relation up to $\rho=0.26\,$g/cm$^3$.

\subsubsection{(b) Frequency shift}

\begin{figure}
\centering
\includegraphics[width=\columnwidth]{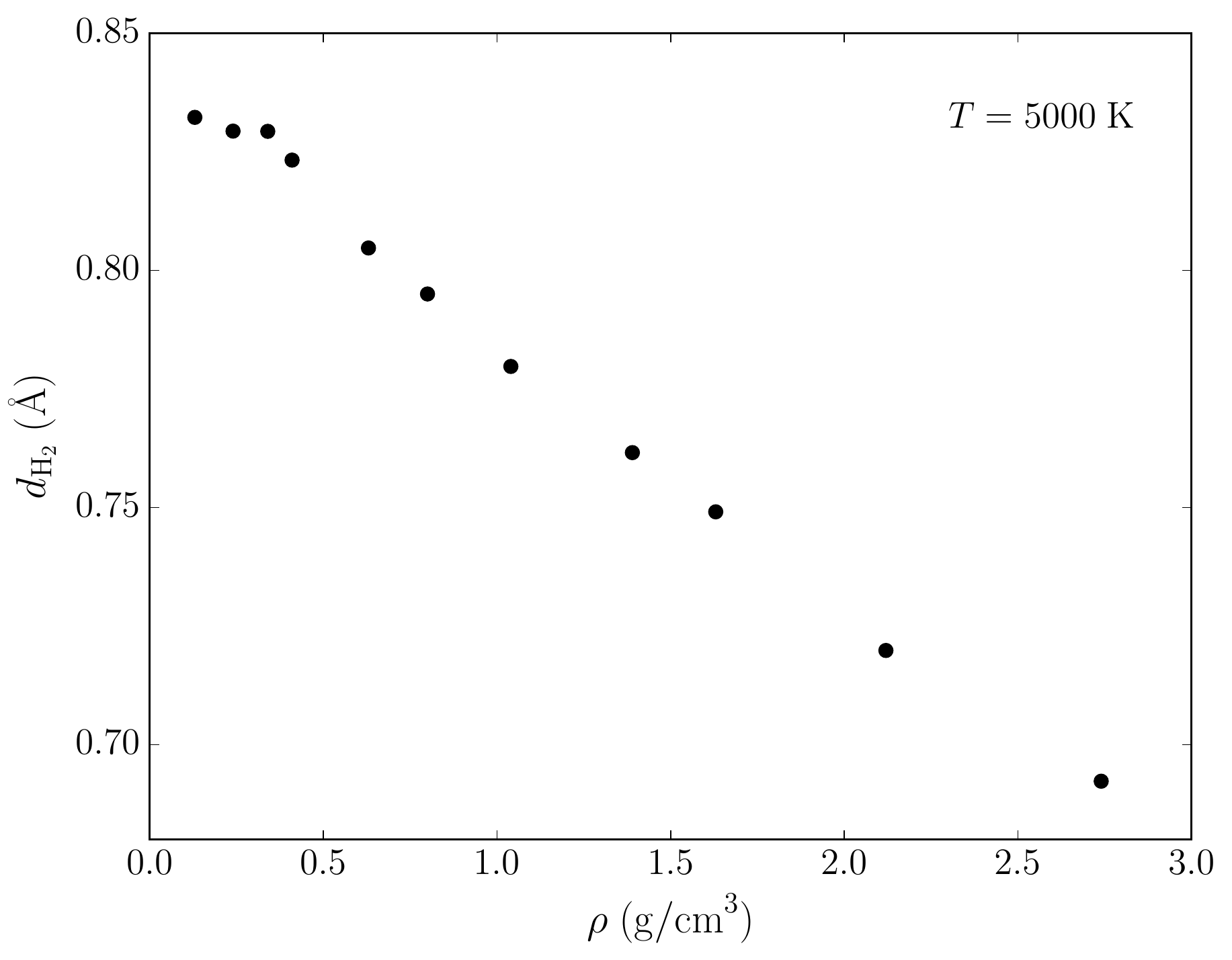}
\caption{Average separation between the two hydrogen atoms for simulations performed at different densities and at $T=5000\,{\rm K}$.}
\label{fig:h2sep}
\end{figure}

The density sequence illustrated in Figure \ref{fig:dens_effect} indicates a systematic shift of the absorption profile towards higher frequencies. This includes the shift of the position of the roto-translational band from $\approx 1000\rm\,cm^{-1}$ in the dilute limit to $>2000\rm \,cm^{-1}$ at $\rho=2.8\,{\rm g/cm}^3$. In the case of rotational transitions, the absorption frequencies (energies) are determined by the rotational constant $B$, which for a $\rm H_2$ molecule is given by,
\begin{equation}
B=\frac{h}{8\pi^2cI}=\frac{h}{4\pi^2cm_{\rm H}d_{\rm H_2}^2} \label{EQB},
\end{equation}
where $h$ is the Planck constant, $m_{\rm H}$ is the mass of the hydrogen atom, $d_{\rm H_2}$ is the distance between the hydrogen atoms in $\rm H_2$ and $I=0.5m_{\rm H}d_{\rm H_2}^2$ is the moment of inertia of $\rm H_2$. In Figure \ref{fig:h2sep} we show the average $d_{\rm H_2}$ values measured along the simulation trajectories at different densities. 
We note that the average separation plotted in Figure \ref{fig:h2sep} is not equivalent to the equilibrium separation. This is because at high temperatures, the anharmonicity of the $\rm H-H$ potential, which is softer for longer interatomic distances, results in an average separation that is larger than the equilibrium bond length \citep{kittel}. The equilibrium $d_{\rm H_2}$ computed using our DFT setup is 0.74{\,\AA} and equals the experimental value \citep[0.74\,\AA,][]{huberDiatomic}.
At $\rho=1.4\,{\rm g/cm}^3$, $d_{\rm H_2}$ decreases by $\approx 10\%$ with respect to the low-density values, which, according to Equation \ref{EQB}, should lead to the increase of the absorption frequency by $\approx 1.23$. This is consistent with the shift of the roto-translational band observed in the simulated profiles. We thus attribute the observed frequency shift of the absorption profiles to the decrease of the $\rm H-H$ bond length induced by dense helium. The frequency shift of the fundamental vibrational band requires a more detailed discussion because of the more complex distortion pattern.

\subsubsection{(c) Distortion of the fundamental band}
\label{sec:dip}

The simulated profiles show a splitting of the fundamental vibrational band and appearance of an intermediate dip. Theses effects have not been considered in previous calculations, but such a density-induced split of the vibrational peak into  $Q_P$ and $Q_R$ branches has been experimentally observed by \citet{bouanich1990collision,chisholm1954induced,crawford1950infra,hare1958pressure}. It is interpreted as an intercollisional interference effect resulting from a time correlation in the intermolecular forces \citep{lewis1976theory}.
The negative correlation resulting in the destructive interference and appearance of the dip results from the opposite direction of dipole moments induced in two successive collisions \citep{kranendonk1968intercollisional,kranendonk1980intermolecular}. Moreover, there are experimental indications of a density-induced shift of the dip position, which  is estimated at $37\rho \, \rm({g/cm}^3) \rm \,cm^{-1} $ \citep{lewis1976theory}.

\begin{figure}
\centering
\includegraphics[width=\columnwidth]{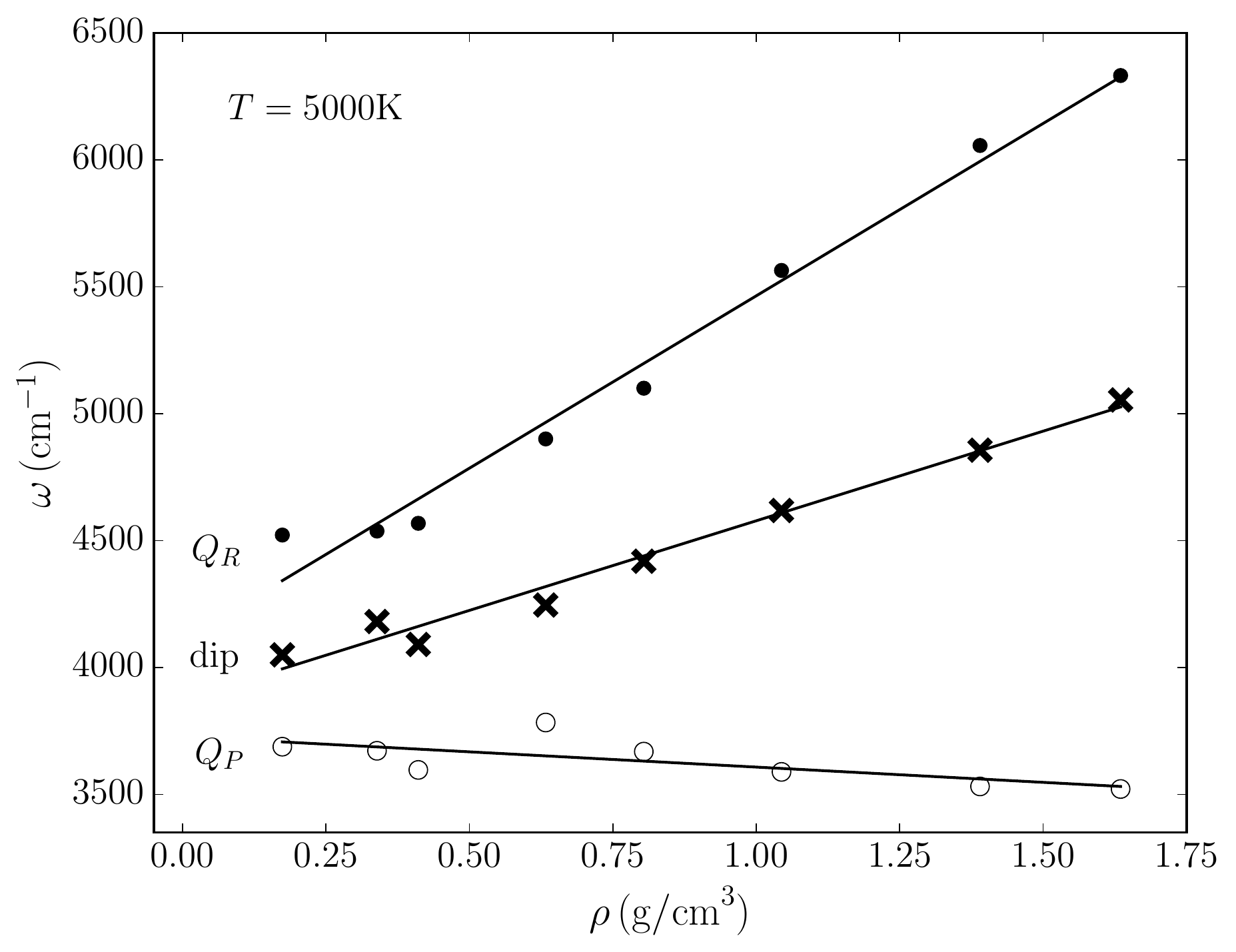}
\caption{Position of the fundamental band interference dip (crosses), $Q_P$ branch (open circles) and $Q_R$ branch (full circles) for various helium densities, at $T=5000$\,K. The data points were extracted from our simulation results and the lines are linear fits.}
\label{fig:peakshift}
\end{figure}

The density dependence of the position of $Q_P$ and $Q_R$ branches and the dip are given in Figure \ref{fig:peakshift}. The positions of the $Q_R$ branch and the dip show a linear dependence on density, while the $Q_P$ component is almost fixed. We note that the shift of the $Q_R$ branch is about twice as big as the shift of the dip. This is an interesting observation that requires further analysis.

\begin{figure}
\centering
\includegraphics[width=\columnwidth]{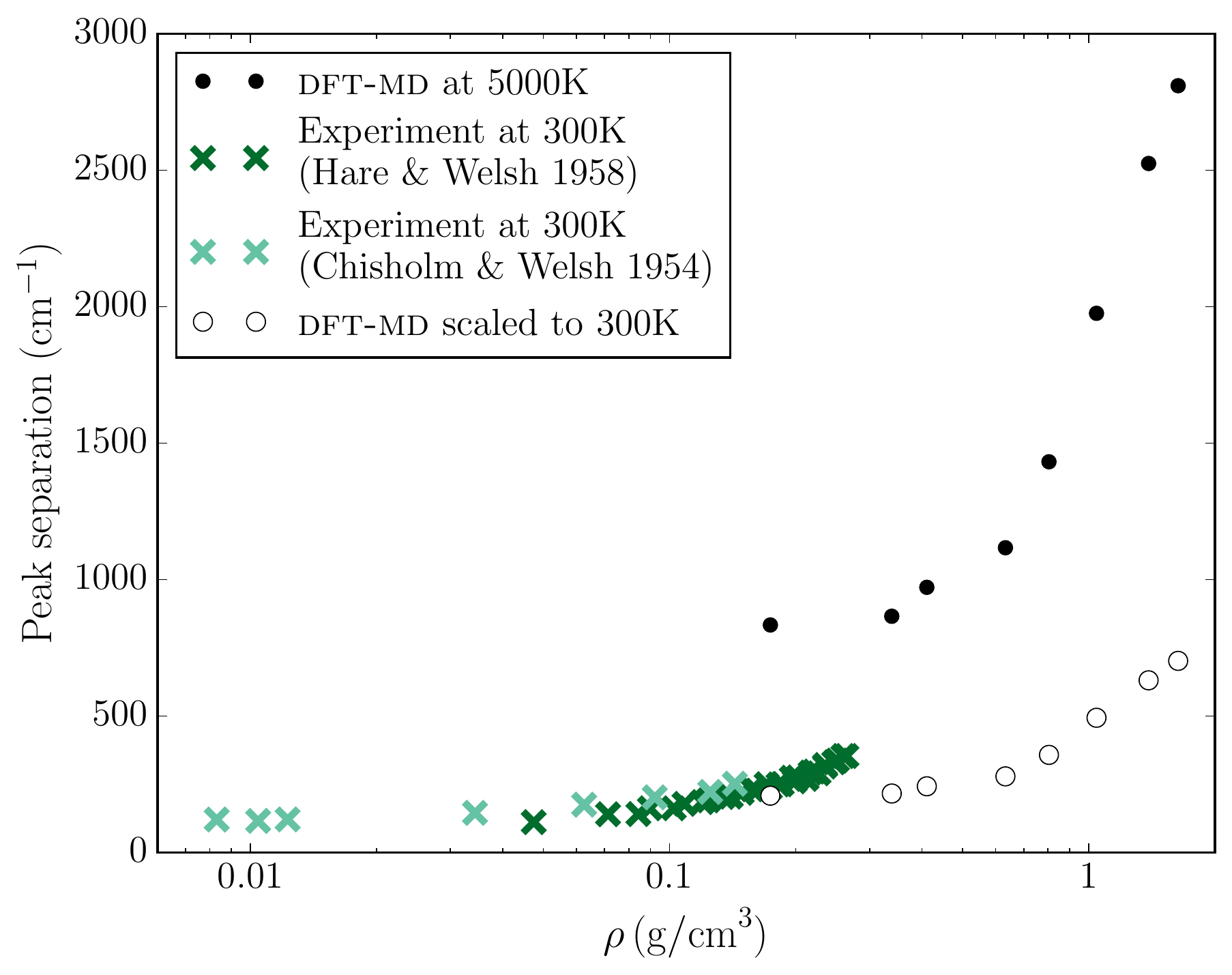}
\caption{Separation between the peaks on either sides of the fundamental band intercollisional dip (distance between the $Q_P$ and $Q_R$ branches). Full circles show the results of our simulations at 5000\,K, open circles are our results results rescaled to 300\,K (divided by a $\sqrt{5000/300}$ factor to account for the temperature difference between simulations and experiments), and crosses indicate the measurements of \cite{chisholm1954induced} and \cite{hare1958pressure}, performed at room temperature.}
\label{fig:peak_sep} 
\end{figure}

We first discuss the size of the splitting into the $Q_P$ and $Q_R$ branches. Assuming that the splitting is caused by the interference effect, we compared the separation observed in our simulated spectra (measured as the distance between the peaks on either side of the dip) to the values reported in experimental studies \citep{chisholm1954induced,hare1958pressure}. The comparison is shown in Figure \ref{fig:peak_sep}. We observe that the separation increases with density (both for the simulated and experimental data) and that at a given density simulations yield a significantly higher peak separation.

The increase of the separation between the $Q_P$ and $Q_R$ peaks can be understood from previous theoretical considerations. Indeed, \cite{kranendonk1968intercollisional} demonstrated that the peak separation should be proportional to the frequency of H$_2$-He collisions, which increases with density. Moreover, this collision frequency dependence might explain the mismatch between the experimental and simulated data as both sets of results were produced at significantly different temperature regimes. In the kinetic theory, the collision frequency is proportional to the mean particle speed $\bar{v}$, which depends on the temperature as,
\begin{equation}
\bar{v} = \sqrt{\frac{8 k_B T}{\pi m}}.
\end{equation}
Since the discussed simulations were performed at 5000\,K and the experimental data was collected at ambient conditions ($T=300$\,K), the collision frequency in the simulations is expected to be $(5000/300)^{1/2} \approx 4$ times higher than in the experiments. When we rescaled the simulation results by this factor, we found a good agreement with the experimental data, as shown by the open circles in Figure \ref{fig:peak_sep}. This match with the experimental data further validates our simulation results.

We note that for a more straightforward comparison with the experimental data we should perform simulations assuming room temperature. Unfortunately, such low-temperature simulations are problematic because of slower dynamics and of the quantum effects on atomic motions that become more pronounced. As discussed in Section \ref{sec:quantum}, these quantum effects are only approximately accounted by our simulation technique. Eventually, a more sophisticated simulation approach such as path integral molecular dynamics would give reliable results, but theses simulations are currently computationally too intensive.

Figure \ref{fig:peakshift} indicates that the dip position is shifted linearly. Previous experimental studies have also found such a linear shift in density \citep{bouanich1990collision,mactaggart1971thesis,mckellar1975studies}, but the experimental estimate of the shift \citep[$37\rho \, \rm({g/cm}^3) \rm \,cm^{-1} $,][]{lewis1976theory} is smaller than the simulated value by a factor of 10. As in the case of the separation between the $Q_R$ and $Q_P$ branches, this could be the effect of the much higher simulated temperatures. The current theoretical explanation for this shift is that the H$_2$ dipole moment shifts during collisions \citep{lewis1976theory,lewis1980intermolecular}. These intracollisional shifts are produced in such a way that the destructive interference is maximal at a frequency that is a function of the perturber density. This also implies an asymmetry between the low-frequency and the high-frequency wings of the dip \citep{lewis1983intercollisional}. This asymmetry is possibly visible in our simulations (Figure \ref{fig:dens_effect}) and in experimental measurements \citep{kelley1984asymmetry}.

We notice that for $\rho<0.4\,$g/cm$^3$ the shift of the dip visible in the simulated spectra is barely detectable. Therefore, we cannot exclude the possibility that up to these densities it is much smaller than at higher densities. Interestingly, the progress of the shift of the dip as a function of density resembles the shift of the roto-translational band. Since we explained the latter by the shortening of the separation of hydrogen atoms in H$_2$, it is likely that the change in the rotation of the $\rm H_2$ molecule contributes to the shift of the dip. This is plausible since at large densities ($\rho>0.4$\,g/cm$^3$) the collisional frequency becomes comparable to the rotational frequency. The interplay of these high-density effects illustrates the complexity of the behavior of matter under fluid-like densities.

Both, theoretical studies \citep{kranendonk1968intercollisional} and room-temperature experiments \citep{hare1958pressure} show that the fundamental band dip extends all the way down to $\alpha(\omega_{\rm dip}) = 0$. This is clearly not the case for our simulated spectra (see Figure \ref{fig:dens_effect}). However, since previous studies were focusing on ambient or low-temperature conditions, little is known about the formation of intercollisional interference at high temperatures. It is highly probable that the simulated dip is shallower because of the important broadening arising under high-temperature conditions.

We note that the intercollisional interference could also explain the density-induced flux suppression of the roto-translational band at $\omega \approx 0\,{\rm cm}^{-1}$ observed in Figure \ref{fig:dens_effect}. Theoretical arguments \citep{kranendonk1968intercollisional} suggest the presence of another intercollisional dip at $\omega = 0$. Just like the dip of the fundamental band, this dip should become wider with increasing density and gradually shift the maximum of the roto-translational band towards higher frequencies. This feature has been experimentally observed by \cite{cunsolo1972collision}, who measured a 10\,cm$^{-1}$ wide dip for a pure H$_2$ gas at $\rho=0.01\,{\rm g/cm}^3$ and $T=300\,$K. The simulated H$_2$-He CIA profiles indicate a $\approx 1000\,{\rm cm}^{-1}$ dip at $\rho \approx 1\,{\rm g/cm}^3$, which is thus consistent with the measurements for the H$_2$ gas. We note that very few measurements exist because of the experimental challenges associated with the low frequencies involved \citep{buontempo1975far}. 

To conclude, the distortion of the fundamental band observed in the simulated absorption profiles is consistent with previous measurements and theoretical considerations. With this analysis of the three main distortion effects, we are able to proceed with the formulation of a general distortion model (Section \ref{sec:modelcia}). Before that, we briefly discuss the temperature dependence of the simulated absorption profiles.

\subsection{Temperature dependence}
\label{sec:temperature}

\begin{figure}
\centering
\includegraphics[width=\columnwidth]{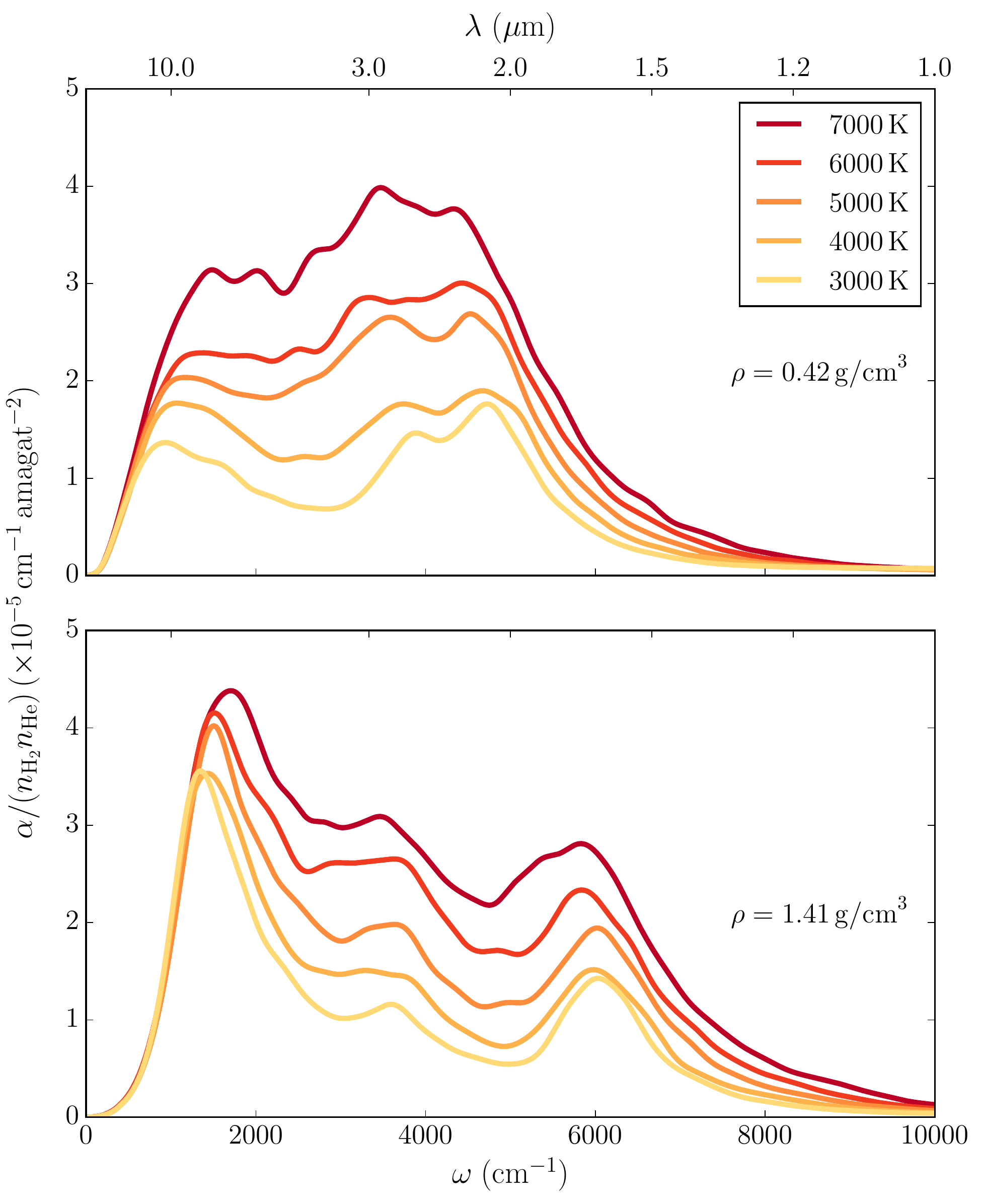}
\caption{DFT-MD H$_2$-He CIA profiles for 5 different temperatures (see legend) and for a density of 0.42\,g/cm$^3$ (top panel) and 1.41\,g/cm$^3$ (bottom panel).}
\label{fig:tempdep}
\end{figure}

So far, our analysis of the distortion effects was limited to the fixed temperature of $T=5000$\,K. To correctly set up the model of high-density distortion of CIA, we checked if the simulations correctly capture the temperature dependence of the absorption profiles. Figure \ref{fig:tempdep} shows the temperature dependence of our simulated profiles for two different helium densities. It is clear that the simulated CIA becomes stronger with increasing temperature and the roto-translational and fundamental bands are more broadened. These two observations are consistent with previous studies (e.g., the same behavior can also be observed in \citealt{abel2012infrared}, see Figure \ref{fig:CIAabel}). In addition to these two well-known effects, Figure \ref{fig:tempdep} shows that the separation of the $Q_P$ and $Q_R$ branches also increases with temperature, which is also expected from the mechanism driving the intercollisional interference (see discussion in Section \ref{sec:dip}). We also note that at the highest reported temperatures the maximum of the roto-translational band is shifted towards higher frequencies, which is also visible in the results of \citealt{abel2012infrared} (Figure \ref{fig:CIAabel}). We took all these effects into account in the construction of the distortion model that is discussed in the next section.

\section{Model of the high-density distortion of H$_2$-H\lowercase{e} CIA}
\label{sec:modelcia}

Using our results, we designed an analytical model of the high-density distortion of the CIA profiles. The idea is that this model can be applied on top of the more accurate dilute-limit ($\rho \rightarrow 0$) calculations, and thus easily implemented in existing white dwarf atmosphere codes.

\begin{figure*}[!tbp]
\begin{centering}
\includegraphics[width=\textwidth]{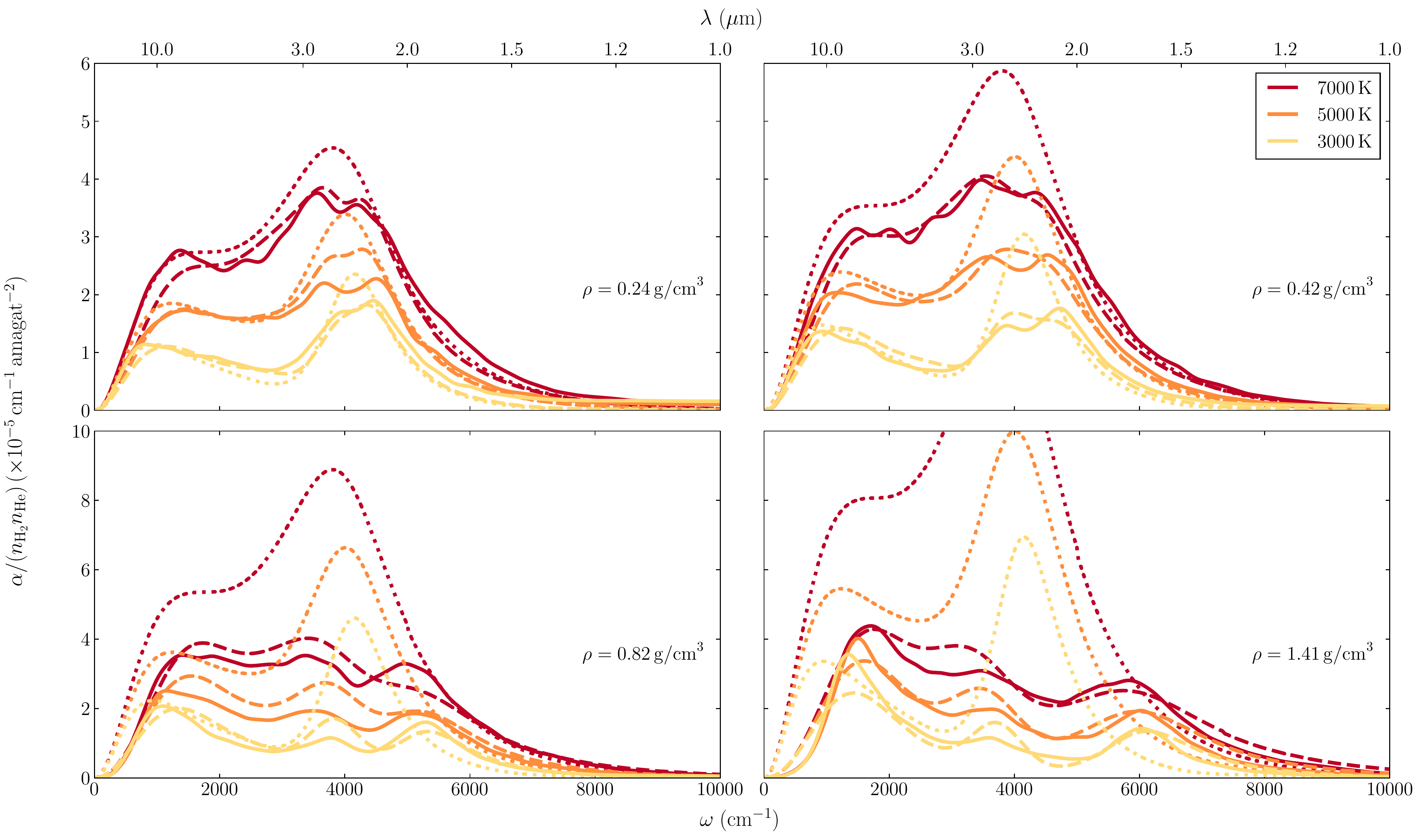}
\end{centering}
\caption{Comparison between our absorption profiles (solid lines), the spectra obtained from our analytical distortion model (dashed lines, Section \ref{sec:modelcia}) and the spectra of \cite{abel2012infrared} scaled for many-body interactions following the method described in Section \ref{sec:previous} (dotted lines). The scale of the vertical axis is not the same for the top and the bottom plots.}
\label{fig:megaplot}
\end{figure*}

Given a low-density limit absorption coefficient $\alpha(\omega,T)$ in $\mathrm{cm}^{-1}\mathrm{amagat}^{-2}$ \citep[e.g.,][]{abel2012infrared}, $\rho$ in g/cm$^3$ and $T$ in K, we model a distorted profile $\alpha'(\omega,T,\rho)$ as,
\begin{equation}
\begin{split}
\alpha'(\omega,T,\rho) = &\left[ \alpha(\omega',T) \, \mathrm{rot}(\omega',T,\rho) + \mathrm{fun}(\omega',T,\rho)\right] \\&\times \mathrm{dip}(\omega',T,\rho).
\end{split}
\end{equation}
The shifted frequency, $\omega'$, reproduces the shift of absorption profiles towards higher frequencies,
\begin{equation}
\omega' = (1 + 0.23 \rho) \omega.
\end{equation}
The linear fit found in Figure \ref{fig:peakshift} was used to implement the density dependence of $\omega'$. The function $\mathrm{rot}(\omega,T,\rho)$ accounts for the enhancement of the roto-translational band with increasing density,
\begin{equation}
\begin{split}
\mathrm{rot}(\omega,T,\rho) = &1 + 3.15 T_{3} \rho g(\omega,1500,800),
\end{split}
\end{equation}
where $T_{3} = T/ 3000\,$K and $g(\omega,\mu,\sigma)$ is the Gaussian function,
\begin{equation}
g(\omega,\mu,\sigma) = \exp \left[ -\frac{(\omega - \mu)^2}{2 \sigma^2} \right].
\end{equation}
The density and temperature dependences of $\mathrm{rot}(\omega,T,\rho)$ were fitted to obtain the right intensity of the roto-translational band. The function $\mathrm{dip}(\omega,T,\rho)$ mimics the splitting of the fundamental band,
\begin{equation}
\mathrm{dip}(\omega,T,\rho) = 1 - \rho e^{-\rho} g(\omega,4100-50 T_{3},800 \rho).
\label{eq:dip}
\end{equation}
The temperature dependence of the center of the Gaussian function in Equation \ref{eq:dip} takes into account the slight shift of the fundamental vibrational band intercollisional dip with respect to temperature (see lower panel of Figure \ref{fig:tempdep}). The density dependence of the width of the Gaussian function accounts for the increased peak separation at higher density (see Figure \ref{fig:peak_sep}). Finally, the $\mathrm{fun}(\omega,T,\rho)$ function accounts for the shift of the fundamental band and the gradual disappearance of the minimum between the roto-translational and fundamental bands,
\begin{equation}
\begin{split}
\mathrm{fun}&(\omega,T,\rho) = 5T_3 ^{0.5} g(\omega,4000-200 \rho T_{3},400 T_{3}^{0.5}) \times \\ & 10^{-5}  \rho e^{-1.2 \rho} + 2 T_3g(\omega,2800,500) \times 10^{-5}  \rho e^{-2 \rho}.
\end{split}
\label{eq:fun}
\end{equation}
Equation \ref{eq:fun} was found by visually fitting the simulated spectra for all  $(T,\rho)$ conditions considered in this work. Figure \ref{fig:megaplot} compares the simulated CIA profiles, the spectra of \citealt{abel2012infrared} (scaled to account for three-body interactions, as discussed in Section \ref{sec:previous}) and the spectra given by our analytical model. Overall, the model reproduces well the main distortion effects described in the previous sections. We note that some arbitrary choices were made in the design of the distortion model, but our goal was to derive an analytical model that reproduces our simulated profiles and which can be applied on top of the H$_2$-He CIA profiles used in white dwarf atmosphere codes. Certainly, a better understanding of the distortion -- obtained either through future experimental studies or more accurate simulations -- may result in a simpler and more physically grounded model.

\section{Implications for the modeling of the atmosphere of cool white dwarf stars}
\label{sec:models}

Our simulations have revealed important high-density distortions of the strength and the shape of H$_2$-He CIA profiles. Using the model described in Section \ref{sec:modelcia}, we implemented these distortions in our atmosphere code to investigate how they affect the synthetic spectra of cool, helium-rich white dwarf stars.

\subsection{Description of the atmosphere code}

The local thermodynamic equilibrium (LTE) atmosphere model code we use is based on the one described in \cite{bergeron1995new}, with the improvements discussed in \cite{tremblay2009spectroscopic} and \cite{bergeron2011comprehensive}. It was further modified to include several non-ideal effects required to properly describe the dense atmosphere of cool helium-rich white dwarf stars. In particular, it includes the non-ideal H$_2$ dissociation equilibrium described by \cite{kowalski2006dissociation}, the non-ideal chemical potentials of helium reported by \cite{kowalski2007equation}, the He-He-He CIA opacities found by \cite{kowalski2014infrared}, and high-density corrections to the free-free absorption of He$^{-}$ and to Rayleigh scattering \citep{iglesias2002density}. Moreover, the energy density and the total number density of each atmospheric layer are computed using the H-REOS.3 and He-REOS.3 ab initio equations of state for hydrogen and helium \citep{becker2014ab}. For mixed H/He compositions, we use the additive volume rule to compute the mass density $\rho(P,T)$ and the internal energy density $u(P,T)$,
\begin{eqnarray}
\frac{1}{\rho_{\mathrm{mix}}(P,T)} &=& \frac{X}{\rho_{\mathrm{H}}(P,T)} + \frac{Y}{\rho_{\mathrm{He}}(P,T)}, \label{eq:rhomix} \\
u_{\mathrm{mix}}(P,T) &=& X u_{\mathrm{H}}(P,T) + Y u_{\mathrm{He}}(P,T), \label{eq:umix} \end{eqnarray}
where $X$ and $Y$ are respectively the mass fractions of hydrogen and helium. Although it does not represent an exact treatment of H/He mixtures, \cite{becker2014ab} showed that using Equations \ref{eq:rhomix} and \ref{eq:umix} yields values that are in good agreement with the real mixture results of \cite{militzer2013ab}, obtained through DFT-MD calculations.

\subsection{Synthetic spectra}
\begin{figure*}
\centering
\includegraphics[width=\textwidth]{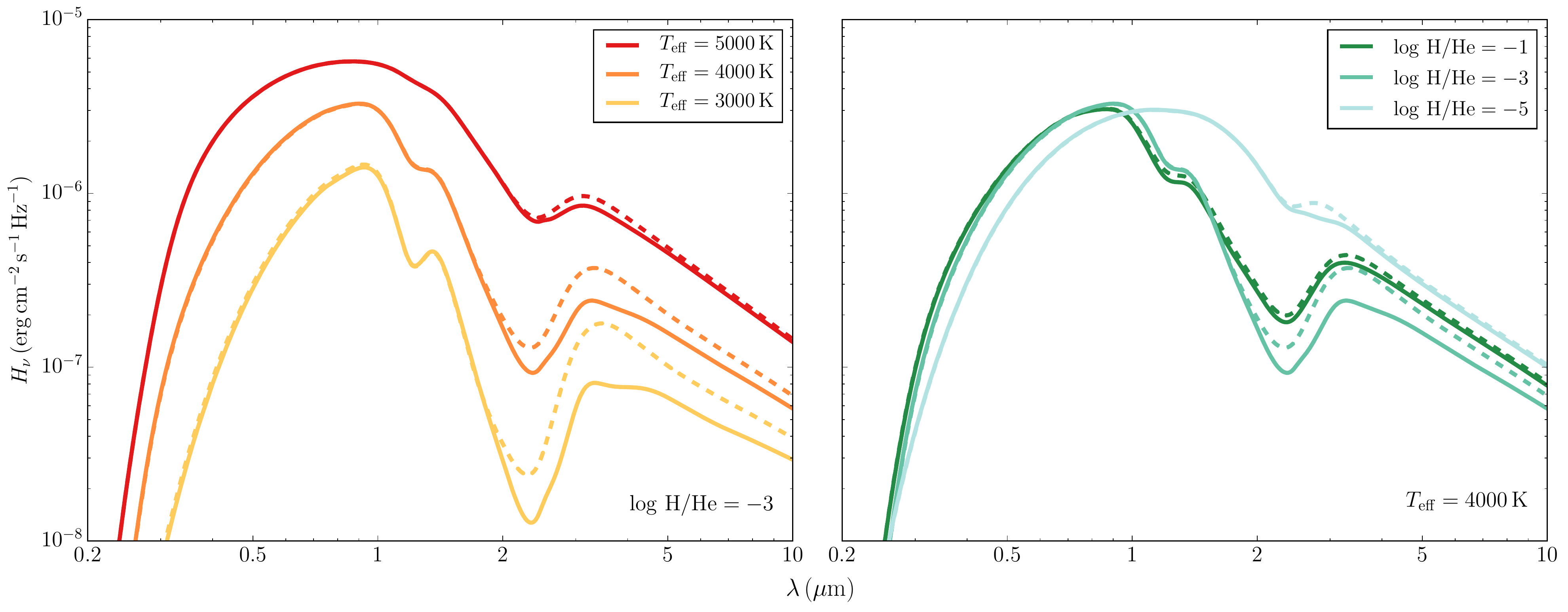}
\caption{Synthetic spectra computed with our distorted H$_2$-He CIA spectra (solid lines) and with the spectra of \citealt{abel2012infrared} (dashed lines). The left plot shows spectra obtained with $\log {\rm H/He}=-3$ and various effective temperatures, while the right plot shows results obtained with various hydrogen abundances and a fix $T_{\rm eff}=4000\,{\rm K}$ effective temperature. All spectra were computed assuming $\log g =8$.}
\label{fig:synspec}
\end{figure*}

Figure \ref{fig:synspec} compares the synthetic spectra computed with the new H$_2$-He CIA profiles to those computed with the spectra of \cite{abel2012infrared}, for various $T_{\rm eff}$ and hydrogen abundances. Accounting for high-density effects in the H$_2$-He CIA results in two important changes. First, a flux depletion redward of 2$\,\mu$m is observed. This effect is a direct consequence of the many-body collisions that arise at high density and cause an important CIA intensity gain. Secondly, we notice a slight distortion of the absorption band at 2.3$\,\mu$m. This distortion is the result of the fundamental H$_2$-He CIA band splitting and shifting. As the absorption bands observed in the synthetic spectra are the result of the sum of the contributions of different atmospheric layers (which have different temperatures and densities), the splitting of the fundamental band is not directly visible.

\begin{figure}
\centering
\includegraphics[width=\columnwidth]{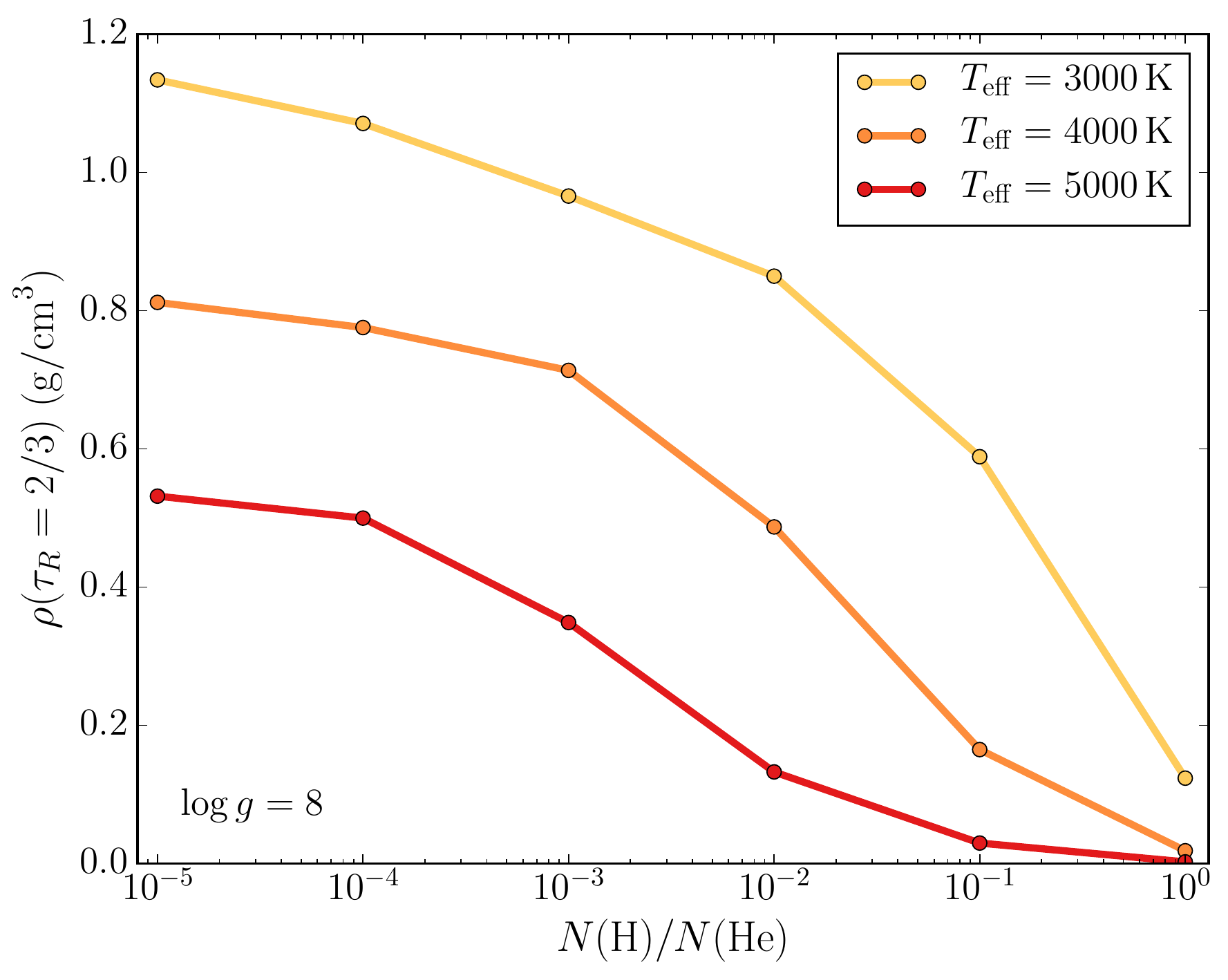}
\caption{Density of our atmosphere 
models at a Rosseland optical depth $\tau_R = 2/3$, as a function of $T_{\rm eff}$ and the hydrogen abundance. These calculations were made assuming $\log g = 8$.}
\label{fig:photodens}
\end{figure}

Figure \ref{fig:synspec} shows that high-pressure CIA distortion effects are more important in the cooler atmospheres. This trend is a direct consequence of the higher photospheric density of low-$T_{\rm eff}$ models. When the temperature is low, there are fewer electrons, which results in a reduction of He$^-$ free-free absorption and an increased transparency of the atmosphere. This raises the density at the photosphere of the star (see Figure \ref{fig:photodens}), which leads to a stronger distortion of CIA.

Figure \ref{fig:synspec} also shows that the discrepancies between the new and old synthetic spectra are mostly important at intermediate hydrogen abundances (around $\log\,{\rm H/He} = -3$). This is the result of the competition between two mechanisms. On one hand, a high hydrogen abundance results in stronger H$_2$-He CIA features, since there is more H$_2$ in the atmosphere. On the other hand, if the atmosphere contains too much molecular hydrogen, it is more opaque and the photosphere is less dense (see Figure \ref{fig:photodens}). This explains why the distortion of CIA is stronger at $\log\,{\rm H/He} = -3$ than at $\log\,{\rm H/He} = -1$ (where the density is too low to produce significant distortion effects) or $\log\,{\rm H/He} = -5$ (where there is too little H$_2$ to produce significant H$_2$-He CIA features).

In general, the density-distorted CIA profiles can contribute to our understanding of the SED of some puzzling cool, helium-rich white dwarf stars. In particular, they might help explain the flux minimum observed in the $3-7\,\mu{\rm m}$ region of the spectrum of \mbox{LHS 3250}, as opposite to the minimum at $2.3 \mu$m predicted by the models \citep[see the comparisons between \textit{Spitzer} photometry and atmosphere models in][]{kilic2009spitzer,kowalski2013cool}. In fact, our new CIA profiles at high densities lead to a significant decrease of the absorption at the position of the fundamental band and an enhancement of the roto-translational band ($3-10\mu$m). This could explain the spectrum of LHS 3250 if the photospheric density for this star is sufficiently high ($\rho>\rm 1.5\,{\rm g/cm}^3$). 

The spectra shown in Figure \ref{fig:synspec} indicate that the distortion effects are not strong enough (or the photospheric density is too low) to improve the fit of this particular star. However, we note that this discrepancy may be caused by a too strong pressure-induced ionization of helium by the chemical equilibrium model implemented in the code. Namely, a slightly reduced ionization would produce less free electrons and thus potentially increase the photospheric density so the CIA distortion could better match the observed spectra of cool, helium-rich stars. Indeed, when computing a model with $T_{\rm eff}=4000\,{\rm K}$, $\log\,{\rm H/He} = -3$ and $\log g = 8$ using the non-ideal helium ionization equilibrium of \cite{kowalski2007equation}, we find a photospheric density of 0.7$\,{\rm g/cm}^3$. For the same atmospheric parameters, if we use the ideal Saha equation to compute the helium ionization equilibrium, the density at the photosphere reaches 1.5$\,{\rm g/cm}^3$. Since the density at the photosphere depends strongly on helium ionization equilibrium, a slight change in it might result in major changes to the signature of H$_2$-He CIA in the synthetic spectra. 

Furthermore, the ionization equilibrium of helium also depends on the applied EOS \citep{bergeron1995new}. This indicates that all problems related to the modeling of these extreme atmospheres must be solved to obtain a satisfactory and final fit to the entire SED of cool, helium-rich stars.

\section{Conclusion}
\label{sec:conclusion}

We applied the ab initio molecular dynamics method to simulate the distortion of H$_2$-He CIA at high densities and temperatures that resemble the physical conditions found at the photospheres of cool, helium-rich white dwarf stars. At low densities the obtained absorption profiles are consistent with previous calculations and experimental data. However, under high-density conditions ($\rho>0.1\,{\rm g/cm}^3$), we found that the H$_2$-He CIA absorption becomes significantly distorted in a way that is not accounted by  current models, but that is consistent with experimental findings. For densities beyond $0.1\,{\rm g/cm}^3$, the integrated absorption profile increases because of multi-atomic collisions, the absorption is shifted towards higher frequencies and the fundamental vibrational band is split. We provided a detailed analysis of these phenomena and constructed an analytical model of the distortion that can be easily applied on top of any low-density H$_2$-He CIA absorption profiles used in current white dwarfs atmosphere codes.

The density-driven evolution of the distortion of H$_2$-He CIA changes the maximum of the absorption from $2.3\,\mu$m to $3-7\,\mu$m, through the enhancement of the roto-transitional absorption peak and the depletion of the vibrational band. This behavior seems to be consistent with the spectra of some cool, helium rich stars, including LHS 3250. The simulations of the representative white dwarf spectra indicate that these distortion effects can significantly affect their mid-IR spectra. They are particularly important for objects cooler than $T_{\rm eff}=4000\,{\rm K}$ with mixed H/He atmospheres. However, the strength of the distortion depends on the photospheric density, which is governed by the still highly uncertain ionization equilibrium of helium.

While the new high-density H$_2$-He CIA distorted spectra might be part of the solution to explain the peculiar SED of some cool white dwarf stars, further effort on the constitutive physics of cool, helium-rich white dwarf atmospheres is needed before we can obtain reliable spectral fits for these objects. In particular, the helium ionization equilibrium, which determines the photospheric density and thus the strength of the CIA profile distortion, must be revisited.

\acknowledgments
This work was supported in part by NSERC (Canada). S.B. acknowledges financial support from DAAD (Germany) for his research stay at Forschungszentrum J\"ulich GmbH.

\bibliographystyle{aasjournal}
\bibliography{references}

\end{document}